\documentclass[a4paper,11pt]{article}
\usepackage{pos}
\usepackage{tikz}

\usetikzlibrary{arrows,snakes,shapes.arrows,decorations.markings}
     \tikzset{>=triangle 90}
     \tikzstyle{gr}=[draw,circle,green!50!black,fill=green!50!black,scale=.6]
     \tikzstyle{Bl}=[draw,circle,blue,scale=.6]
     \tikzstyle{R}=[draw,circle,fill=red,scale=.6]
     \tikzstyle{bl}=[draw,circle,fill=black,scale=.6]
     \tikzstyle{rc}=[circle,fill=red,scale=.6]

\title{The weak gravity conjecture in perturbative strings}

\author*{Matteo Lotito}

\affiliation{Instituto de F\'isica Te\'orica IFT-UAM/CSIC,\\
  C/ Nicol\'as Cabrera 13-15, Campus de Cantoblanco, 28049 Madrid, Spain}

\emailAdd{matteo.lotito@ift.csic.es}

\abstract{In this note we give a summary of \cite{Heidenreich:2024dmr} in which we proposed a proof of the weak gravity conjecture in perturbative string theory. While the WGC is well established, checked in many examples, and many of the ingredients we use have previously appeared in the literature, a comprehensive proof from the top-down was still missing.
The present work focuses on the bosonic string as a proof of concept, while the generalization to superstring cases  is to appear in a forthcoming paper.
This note is based heavily on \cite{Heidenreich:2024dmr} and on a talk given at the Corfu2025 Workshop on Quantum Gravity and Strings.
}

\FullConference{
Corfu Summer Institute 2025 ``School and Workshops on Elementary Particle Physics and Gravity''\\
27 April - 18 September 2025\\
Corfu, Greece
}

\begin{document}
\renewcommand{\hookAfterAbstract}{
    \par\bigskip\bigskip\bigskip
    Report number: IFT-UAM/CSIC-26-59
}

\maketitle

	\section{Introduction}

A theory of quantum gravity (QGT) is supposed to be a model which captures the features of general relativity and of quantum mechanics in a single consistent framework.
Finding the general structure of such a theory is, naturally, a tall order.

The swampland program \cite{Vafa:2005ui} (see \cite{Palti:2019pca, Brennan:2017rbf} for reviews) is an approach that tries to inform us of at least part of the structure of this putative theory by asking what are the effects that quantum gravity has from the perspective of an effective field theory (EFT) in the low energy limit.
Since its inception about two decades ago, the swampland program has been an extremely active field of research. Within it, various conjectures have been proposed 
\cite{Arkani-Hamed:2006emk, Ooguri:2006in, Ooguri:2018wrx}, these provide constraints on the properties of candidate EFTs to be suitably completed to a consistent UV theory.
The space of EFTs can be divided into a ``landscape'', consisting of theories that are indeed the low energy limit of a quantum gravity theory, and the ``swampland'', which instead contains seemingly consistent EFTs that cannot have a consistent UV completion to a QGT.

The various swampland conjectures originate from different features of the EFTs (or expectations of the QGT) and constrain the landscape in different ways. In talks and reviews they are often organized with an inverse proportionality relating their usefulness (constraining power) versus their mathematical rigor\footnote{by which I mean the rigor at which the average physicist would consider a statement proven}.

In the hope of streamlining and clarifying the picture of allowed QGTs and their associated EFTs, one should work in a dual fashion: on one hand testing current conjectures in interesting examples, as well as formulating modifications of existing or  new conjectures to further constrain the landscape; on the other hand, one should try to exploit techniques and structures of frameworks that we understand well (e.g. 2d CFTs) to try to hone and prove, albeit with the necessary simplifying assumptions, some of these conjectures.

The present work is in the spirit of the latter approach and it provides a proof of the Weak Gravity Conjecture in perturbative string theory.
Although the presentation and the publication it is based on \cite{{Heidenreich:2024dmr}} is for the bosonic string, we believe that the fundamental results will carry over to the superstring case \cite{HL26}.

Note that most swampland conjectures are defined, and in particular tested for, within the context of string theory.
It is an open question, though far beyond the scope of the present work, to ask whether all the EFTs in the landscape come from string theory (equivalently, if any QGT can be described as a string theory). At present, we assume the affirmative, and so we will use QGT and string theory as synonyms, no less for the fact that the string theory framework is the only one where we can test conjectures. Nevertheless, this is an interesting question that deserves further investigation.

\subsection{Weak Gravity Conjecture}
One of the most established conjectures of the swampland program is the so-called Weak Gravity Conjecture (WGC) \cite{Arkani-Hamed:2006emk} (see also a recent review \cite{Harlow:2022ich}).
. This arises from the theoretical (and phenomenological)  requirement that extremal black holes should decay.
The statement of the WGC in its mildest form says that in a QGT with a massless photon, the spectrum should contain a charged state satisfying
\begin{equation}
\dfrac{|q|}{m} \geq \left.\dfrac{|Q|}{M}\right|_{\text{extremal BH}}.
\end{equation}
A state satisfying this condition is called \emph{superextremal}.
The statement compares the spectrum of the QGT in the UV with semiclassical properties of a black hole.
In order to have a good semiclassical description of black hole solutions we need to qualify the RHS in more detail.
We need to compare the spectrum against \emph{large} black holes.
It is interesting to ask how much we can lower the mass scale, but in any event if we went all the way down to the Planck scale we would lose the constraining power of the statement. Conversely, if we knew the spectrum precisely in that regime, presumably we would have control over the full QGT, negating the need for any swampland conjectures.
Furthermore, here we are only considering electrically charged black holes, but extremal electrically charged black hole solutions are singular. Nevertheless, we can construct a family of solutions of a given charge whose limits are these extremal cases. As we can apply the WGC for the entire family, we also extrapolate to these limiting cases. 

The expression above gives the \emph{mild} form of the WGC. Since the original formulation, several refined statements have been proposed.
The \emph{lattice} WGC requires that the statement of the WGC is satisfied in the entire charge lattice, that is, there is a superextremal state for every site of the charge lattice of the theory.
A slightly weaker statement is the \emph{sublattice} WGC \cite{Heidenreich:2016aqi}, which only requires a sublattice of the full charge lattice (a priori not specified) to be filled with superextremal states.\footnote{A similar notion is the \emph{tower} WGC \cite{Andriolo:2018lvp} which requires the presence of superextremal states in every direction of charge space, the difference being that it does not specify, even in principle, the index of a sublattice within the charge lattice.}

A different notion is the \emph{Ooguri-Vafa criterion} for the WGC, introduced in \cite{Ooguri:2016pdq}. This pertains to the analysis of theories with supersymmetry. This essentially means that the WGC bound can only be saturated in a supersymmetric theory.
More precisely, if there is a superextremal state that saturates the bound, then it must be BPS.
Conversely, if there are superextremal states that satisfy but do not saturate the bound, these cannot be BPS.

\medskip

The version of the WGC that we prove in our work is the Ooguri-Vafa version of the sublattice WGC.
In the bosonic string case, this means that we find a sublattice of the charge lattice that contains superextremal states that satisfy but do not saturate the WGC bound.
In the superstring case, yet to be finalized, we might encounter non-saturating superextremal states, just as in the bosonic case, but in addition we might expect saturating states, but in this instance, this will go hand in hand with verifying that they are BPS.

\subsection{Repulsive Force Conjecture}
The mnemonic that ``gravity is the weakest force'' does not describe well the statement of the WGC, since so far we have only discussed the spectrum of a candidate theory.
The connection to the lore is clearer from the related, but in general inequivalent, Repulsive Force Conjecture (RFC)
\cite{Palti:2017elp, Heidenreich:2019zkl}, which constrains the dynamics, precisely imposing a bound on the resultant of the long-range forces. 
The statement of the RFC is that a pair of identical (charged) states is subject to a force
    \begin{equation}\label{RFC}
    F = F_{\text{grav}, \Phi^0} + F_{\text{gauge}} + F_{\text{moduli}} \geq 0.
    \end{equation}
States satisfying this condition are called \emph{self-repulsive}.

    In the absence of moduli (i,e, setting the last factor of \eqref{RFC} to 0) one indeed finds that the statements of the RFC and the WGC are equivalent.
    In general, however, the corrections due to arbitrary moduli are unknown, leading to inequivalent statements.

\section{Why should the WGC be true?}
    
\begin{figure}[htbp]\centering
	\begin{tikzpicture}
		\fill[blue!10] (0,1) -- (1,1) -- (7,2.5) -- (7,5) -- (0,5) -- cycle;
		\draw[->] (0,0) -- (7,0) node[above right] {$Q$}; 
		\draw[->] (0,0) -- (0,5) node[above left] {$M$}; 
		
		\draw[dashed] (1,1) -- (7,2.5) node[below right] {$M_{\text{EXT}}(Q)$};
		\draw[red,dashed]  (0,1) node[left] {$M_{Pl}$}-- (5,1);

			\draw[green!30, densely dotted ] (1.6,2.83) -- (5.3,3.93); 
		\node[green!30] at (3.5,3.4) {$\bullet$};
\node[green!30] at (5.5,4) {$\bullet$};
\node[green!30] at (4.5,3.7) {$\bullet$};
\node[green!30] at (1.5,2.8) {$\bullet$};

\node[orange] at (3,0.75+0.75) {$\bullet$};
\node[orange] at (4,1+0.75) {$\bullet$};
\node[orange] at (5,1.25+0.75) {$\bullet$};
		\draw[orange, densely dashed ] (5,2) -- (5.5,4); 
		\draw[orange, densely dashed ] (4,1.75) -- (4.5,3.7);
		\draw[orange, densely dashed ] (3,1.5) -- (3.5,3.4); 
				
\node[rotate=0] at (6,3) {\small{Hawking Radiation}};
\node[rotate=17] at (2.8,3.5) {\small{Gravitational Collapse ->}};
		\end{tikzpicture}
\caption{Formation of stable black hole remnants}
\label{figstable}
\end{figure}
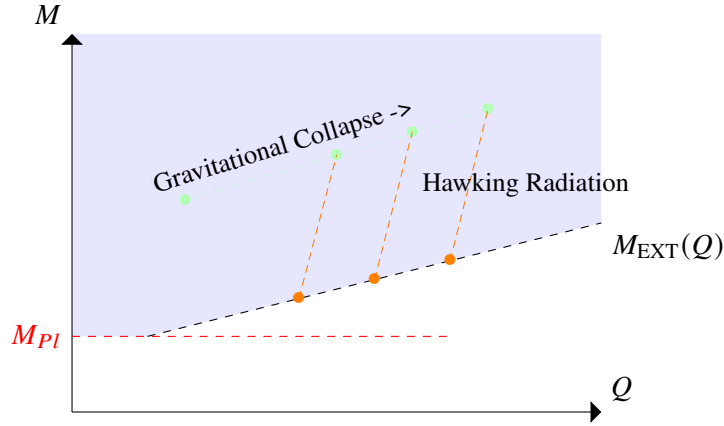

The motivation for the WGC to hold comes from black hole physics arguments and can be summarized with the intuition that extremal black holes should decay.
Gravitational collapse would tend to form more and more massive black holes and even accounting for Hawking radiation (discharging to extremal black holes), we would obtain stable states along the extremality bound, where Hawking radiation is no longer active.
The gauge charge of the effective $U(1)$ theory would look like a global symmetry and the process described above, see figure \ref{figstable}, would produce stable black hole remnants.

\begin{figure}[htbp]\centering
	\begin{tikzpicture}
		\fill[blue!10] (0,1) -- (1,1) -- (7,2.5) -- (7,5) -- (0,5) -- cycle;
		\draw[->] (0,0) -- (7,0) node[above right] {$Q$}; 
		\draw[->] (0,0) -- (0,5) node[above left] {$M$}; 
		
		\draw[dashed] (1,1) -- (7,2.5) node[below right] {$M_{\text{EXT}}(Q)$};
		\draw[red,dashed]  (0,1) node[left] {$M_{Pl}$}-- (5,1);

			\draw[green!30, densely dotted ] (1.6,2.83) -- (5.3,3.93); 
		\node[green!30] at (3.5,3.4) {$\bullet$};
\node[green!30] at (5.5,4) {$\bullet$};
\node[green!30] at (4.5,3.7) {$\bullet$};
\node[green!30] at (1.5,2.8) {$\bullet$};

\node[orange] at (3,0.75+0.75) {$\bullet$};
\node[orange] at (4,1+0.75) {$\bullet$};
\node[orange] at (5,1.25+0.75) {$\bullet$};
		\draw[orange, densely dashed ] (5,2) -- (5.5,4); 
		\draw[orange, densely dashed ] (4,1.75) -- (4.5,3.7);
		\draw[orange, densely dashed ] (3,1.5) -- (3.5,3.4); 

		\draw[black, ->, >=stealth] (5,2) -- (1.5,1.6); 
		\draw[black, ->, >=stealth] (4,1.75) -- (2.1,1.5);
		\draw[black, ->, >=stealth] (3,1.5) -- (1.5,1.2); 

		\draw[purple, ->, >=stealth] (5,2) -- (0.7,0.1); 
		\draw[purple, ->, >=stealth] (4,1.75) -- (1,0.3);
		\draw[purple, ->, >=stealth] (3,1.5) -- (0.3,0.15); 

		\end{tikzpicture}
\caption{Decay of extremal black holes to subextremal black holes (black) due to the presence of  superextremal states (purple) in the spectrum.}
\label{figunstable}
\end{figure}
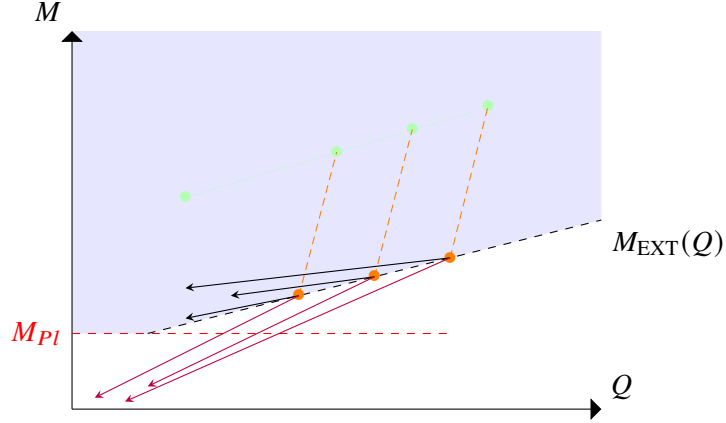

If, on the other hand, the spectrum contains states that allow extremal black holes to decay, there would not be stable remnants, as shown in figure \ref{figunstable}.
In this sense, the WGC is thought to be a consequence of the statement that there are no global symmetries in quantum gravity \cite{Harlow:2018tng}.

\medskip

The WGC has been shown to hold in many string theory examples, in fact strong forms, like the lattice and sublattice versions hold in heterotic theories.
Toroidal compactifications of the heterotic string are particularly useful testing grounds as in these cases one can compute explicitly both sides of the inequality.
While the full lattice WGC holds in simple toroidal compactifications, it fails for for compactifications with orbifolds, e.g. on $T^6/\mathbb{Z}_2$. In this case one can still compute explicitly and finds that the sublattice version of the WGC still holds.
This is generically true for all known examples in which explicit computations can be performed. The goal of our work is to show that this is not simply an artifact of having chosen ``nice'' examples, but rather a feature of a generic string theory.

\section{What goes in our proof?}
Having introduced the statement of the WGC and provided at least a sketch of a motivation for it, we now describe the main ingredients that enter our proof. The key point is that all these ingredients are universal properties of string constructions, thus they do not depend on specific choices of sample models (compactification manifolds, etc.).

There are the main ingredients that we use in our derivation:
\begin{itemize}
    \item modular invariance on the string worldsheet allows us to guarantee the presence of specific charged states in the spectrum \cite{Montero:2016tif}. For these states we know the relation between their mass and charge and they will be our candidate supextremal states;
    \item the map between the EFT and the worldsheet description allows us to express three- and four-point functions as correlators on the worldsheet CFT. Combining these, we can explicitly compute the long-range forces between charged states. This allows us to show that the sublattice version of the RFC holds;
    
    \item the semiclassical description of black hole solution with a two-derivative effective action lets us connect the condition of self-repulsiveness from the RFC, to the superextremality condition of the WGC if the former holds everywhere in moduli space.
\end{itemize}
The explicit derivation pf \cite{Heidenreich:2024dmr}, which we summarize in the following, was for the bosonic string. While we believe an analogous result should hold for type II and  heterotic, we will have to revisit the analysis, as the more complicated structure of the theory might enter for each of the previously sketched steps.
Furthermore, one could complain about the presence of tachyons in bosonic strings. Since this is supposed to be an appetizer for the generalization to superstrings, we will not worry about this point. Furthermore, one may ask whether the WGC should hold as is or should be modified whenever these instabilities are present. The results of our analysis seem to suggest that the WGC indeed still holds in these pathological cases.

We now describe in more detail each of these steps.
 
\subsection{Modular Invariance and Spectral Flow}
The spectrum of a perturbative string theory is encoded in the partition function of its worldsheet CFT on a torus.
This is given by
\begin{equation}
		Z(\tau, \bar{\tau}) = \text{Tr}\Bigl[ q^{L_0 - \frac{c}{24}} \bar{q}^{\tilde{L}_0 - \frac{\tilde{c}}{24}}\Bigr]  = \sum q^{h-\frac{c}{24}} \bar{q}^{\tilde{h}-\frac{\tilde{c}}{24}}
				\, , \quad  q = e^{2\pi i \tau}.
\end{equation}
Modular invariance under $SL(2,\mathbb{Z})$ requires that
\begin{equation}
Z(\tau, \bar{\tau}) = Z(\tau+1,\bar{\tau}+1) = Z\left(-\frac{1}{\tau},-\frac{1}{\bar{\tau}}\right).
\end{equation}
When there are global $U(1)$ symmetries (thus conserved currents) on the worldsheet, one can study the ``flavored'' partition function by introducing chemical potentials associated to those conserved charges, essentially weighing each state in the Hilbert space by the eigenvalue of the zero mode with respect to that current.
For $U(1)$'s we can write 
\begin{equation}
	Z(\mu, \tau; \tilde{\mu}, \bar{\tau}) = \sum q^{h-\frac{c}{24}} \bar{q}^{\tilde{h}-\frac{\tilde{c}}{24}} y^{Q}\tilde{y}^{\tilde{Q}}
\end{equation}
with $y^{Q} = e^{2\pi i \mu_a Q^a},  \tilde{y}^{\tilde{Q}} = e^{-2\pi i \tilde{\mu}_{\tilde{a}} \tilde{Q}^{\tilde{a}}}$,\\ This expression is no longer modular invariant, though it still transforms in a well defined way:
\begin{equation}\label{flavormodular}	Z(\mu,\tau;\tilde{\mu},\bar{\tau}) =\left\{ \begin{array}{l}
Z(\mu,\tau+1;\tilde{\mu},\bar{\tau}+1) \\[.3em]
 e^{-\dfrac{\pi i}{\tau} \mu^2 +\dfrac{\pi i}{\bar{\tau}} \tilde{\mu}^2} Z\left(\dfrac{\mu}{\tau},-\dfrac{1}{\tau};\dfrac{\tilde{\mu}}{\bar{\tau}},-\dfrac{1}{\bar{\tau}}\right) \end{array}\right.
\end{equation}
This transformation property appeared in \cite{Benjamin:2016fhe}, but we also proposed an alternative derivation of it in \cite{Heidenreich:2024dmr}.
Together with this, one can use the fact that the line operator on the torus parametrized by the chemical potential, i.e. the insertion giving rise to the $y^{Q}\tilde{y}^{\tilde{Q}}$ contribution in the flavored partition function is invariant under
\begin{equation}\label{period}
(\mu, \bar{\mu})\to (\mu +\rho, \bar{\mu} + \bar{\rho})
\end{equation}
for any $(\rho,\tilde{\rho})$ in the period lattice $\Gamma$, dual to the charge lattice and, in general, a subset of it, $\Gamma \subseteq \Gamma_Q$.
Combining eq. \ref{flavormodular} and eq. \ref{period} we find the quasiperiodicity condition
\begin{equation}
Z(\mu + \tau \rho, \tau, \tilde{\mu} + \bar{\tau} \tilde{\rho}, \bar{\tau}) = e^{-2\pi i \mu \rho - \pi i \rho^2 \tau + 2 \pi i \tilde{\mu}\tilde{\rho} + \pi i \tilde{\rho}^2 \bar\tau} Z(\mu, \tau, \tilde{\mu}, \bar{\tau}) \,, \qquad (\rho,\tilde{\rho}) \in \Gamma \,. \label{eqn:quasiperiod}
\end{equation}
Now we can rewrite the partition function by introducing 
$\widehat{h} = h-\frac{1}{2} Q^2$ and $\widehat{\tilde h} = \tilde{h}-\frac{1}{2} \tilde{Q}^2$
so it reads
\begin{equation}
Z 
=  \sum q^{\widehat{h}-\frac{c}{24}} \bar{q}^{\widehat{\tilde h}-\frac{\tilde{c}}{24}} q^{\frac{1}{2}(Q+\rho)^2} y^{Q+\rho} \bar{q}^{\frac{1}{2} (\tilde{Q}+\tilde{\rho})^2} \tilde{y}^{\tilde{Q}+\tilde{\rho}} \,.
\end{equation}
This finally tells us that the spectrum is invariant under
z\begin{equation}
(Q,\tilde{Q}) \to (Q,\tilde{Q}) + (\rho,\tilde{\rho}) \,,\qquad (\rho,\tilde{\rho}) \in \Gamma \,, \qquad \text{with $\widehat{h}$, $\widehat{\tilde h}$ held fixed.} \label{eqn:SpecFlow}
\end{equation}
Applying this to the identity operator, we can populate all states with $(Q,\tilde Q) \in \Gamma$ and these have conformal dimension $h= \dfrac{1}{2}Q^2, \tilde{h}=  \dfrac{1}{2}\tilde{Q}^2$.

Therefore, the bosonic string contains an infinite tower of states with spectrum
\begin{equation}\label{spectrum}
\frac{\alpha'}{4} m^2 = \frac{1}{2} \max(Q^2, \tilde{Q}^2) -1 \,, \qquad \text{for all $(Q,\tilde{Q}) \in \Gamma$.}
\end{equation}
This shows that these states exist, but we have not yet demonstrated that they are superextremal\footnote{They do so in simple heterotic orbifold examples.}. For this we need to continue with the other ingredients and later combining them.

\subsection{Long-Range Forces}
The second step we have to perform is to compute the self-force between pairs of identical charged states. To do so, we need to map the three- and four-point functions of the EFT into correlators on the worldsheet.

\begin{figure}[tbp]\centering
\includegraphics[width=.7\textwidth]{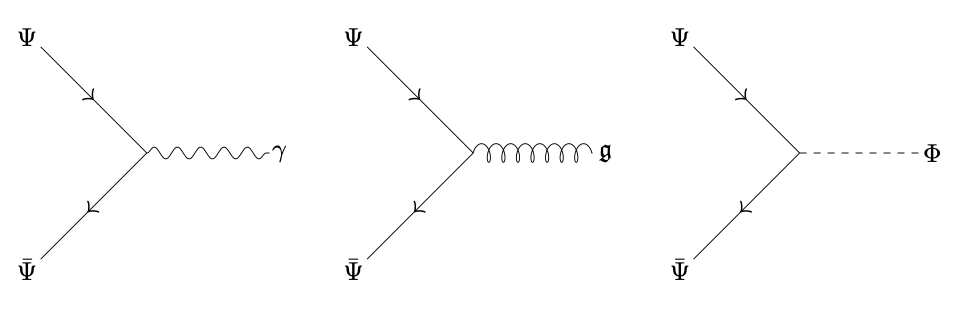}
\caption{Three-point vertices of the charge states with graviton, photon and moduli, respectively that contribute to the long-range forces in the EFT}\label{vertices}
\end{figure}

First, while the physical data is encoded by four-point functions of charged states scattered by gravitons, photons, or moduli, these can be decomposed into combinations of three-point functions and propagators. The latter are then fixed by a choice of normalization, so the physical data is, in essence, encoded in the three-point functions shown in figure \ref{vertices}. Therefore, we want to be able to write $\langle \psi \psi g \rangle, \langle \psi \psi \gamma \rangle, \langle \psi \psi \phi \rangle$ as worldsheet expressions.

To do so, we write the worldsheet CFT as a factorized theory, with an ``external'' CFT that will depend on universal quantities (behaving sort of like normalization factors)\footnote{The external CFT also carries the Lorentz indices, i.e. the spin structure of the states of the EFT.}, and an ``internal'' CFT, which will instead contain the real information that we are after.
Writing the worldsheet theory $\mathcal{T} =  \mathcal{T}_{\text{ext}}\times \mathcal{T}_{\text{int}}$ naturally leads the correlators to also factorize as
\begin{equation}
\langle V_1 V_2 V_3 \rangle = 
\langle V_1^{\mathrm{ext}} V_2^{\mathrm{ext}}
  V_3^{\mathrm{ext}}\rangle_{\mathrm{ext}} \cdot 
  \langle V_1^{\mathrm{int}}
  V_2^{\mathrm{int}} V_3^{\mathrm{int}} \rangle_{\mathrm{int}}.
\end{equation}
Neglecting the external components, we can map the massless states of interest (mediating long-range forces) to internal CFT operators, listed in the following table
\begin{center}
\begin{tabular}{rcc}
	  	state & worldsheet operator & $(h,\tilde{h})$\\\hline
	  	graviton $g_{\mu\nu}$ &1 &(0,0)\\
	  	dilaton $\Phi$ & 1 & (0,0)\\
        photon $A^a_\mu$ & $J^a(z), \tilde{J}^a(\overline{z})$ & (1,0) and (0,1)\\
	  	other massless scalars $\Phi^i$ &$\varphi^i(z,\overline{z})$ & (1,1)
	  \end{tabular}
\end{center}
We can now start computing explicitly the terms that appear in the RFC bound in eq. \ref{RFC}\footnote{In all terms we compute below we remove a common volume prefactor, as the actual force goes as $F/V_{D-2}r^{D-2}$.}.
The first factor is proportional to $\langle\psi \psi g\rangle$, which in the internal CFT reduces to a two-point function $\langle\psi\psi\rangle$\footnote{Abusing notation we will label both the EFT charged states and their internal CFT operator as $\psi$.}
. Normalizing this to 1 and reintroducing the external factors, the overall first term is 
\begin{equation}
F_{\text{grav}, \Phi^0} = -\kappa_d^2 m m'.
\end{equation}

\medskip
The second factor is likewise easily computed. In this case the contribution $F_{\text{gauge}} $ consists of two pieces  proportional to
$\langle\psi \psi J^a(z)\rangle $ and
$\langle\psi \psi \tilde{J}^a(\overline{z})\rangle$.
We can compute this because we know the eigenvalue of the zero mode of the current for these charged states, which implies $\langle\psi J^a(z)\psi\rangle = \frac{i}{z}Q^a\langle\psi\psi\rangle$.
Putting together with the second term ($J\to \tilde{J}$) we get
\begin{equation}
F_{\text{gauge}}=
 \frac{2k_D^2}{\alpha'}\Big( \delta_{ab} Q^a Q^b + \delta_{\tilde{a}\tilde{b}} \tilde{Q}^{\tilde{a}}\tilde{Q}^{\tilde{b}}  \Big). 
\end{equation}

\medskip
The last factor $F_{\text{moduli}}$ is where the most interesting behavior occurs, as this scattering is mediated by arbitrary moduli.
There is still a universal contribution from the gauge part since the operator $\lambda^{a\tilde{b} }(z,\bar{z}) = J^a(z) \tilde{J}^{\tilde{b}}(\bar{z}) $ is a (1,1) primary depending on the currents.
The contribution from this part is computed similarly as $F_{\text{gauge}}$ once deriving \begin{equation}\langle \psi \lambda^{ab}\psi \rangle = \frac{1}{|z|^2} Q^a\tilde{Q}^b.
\end{equation}
Let us denote the other neutral current primaries by $\chi(z,\bar{z})$. These do not mix with $\lambda^{a\tilde{b} }$ (since  $J^a|\chi\rangle= \tilde{J}^a|\chi\rangle=0 \implies \langle \chi \lambda^{a\tilde{b}}\rangle=0$).
So the only term left to compute is
\begin{equation}\label{3pointchi}
\langle \psi \chi(z,\bar{z}) \psi\rangle.
\end{equation}
A quick way to determine this is to decompose the internal CFT \`a la Sugawara \cite{Polchinski}, factoring out the current part.
The stress tensor of the CFT is written as 
$$
T(z) = \hat{T}(z) + T^J(z)
$$ with the current part defined as
$T^J(z)= \dfrac{\delta^{ab}}{2} : J^a(z) J^b(z):$, with similar expressions for the right moving parts.
In the paper \cite{Heidenreich:2024dmr} we spend some time checking that the algebra of Virasoro generators decomposes and behaves appropriately, since we have to introduce insertions of zero modes of the two components of the stress tensor.
For the sake of brevity, in the present context we summarize this by stating that the neutral current moduli and the charged states have the following conformal dimensions under the two factors
\begin{center}\begin{tabular}{rcc}
		Operator &  $(h^J, \tilde{h}^J)$ & $(\widehat{h}, {\widehat{\tilde h}})$\\\hline
		$\psi(z,,\bar{z})$ &$\left( \frac{1}{2}Q^2, \frac{1}{2}\tilde{Q}^2 \right)$ & (0,0)\\
		 $\chi(z,,\bar{z}) $ &$(0,0)$ & (1,1)
	\end{tabular}\end{center} 
It is no coincidence that this reminds us of the shift in conformal dimensions that we performed in the previous section to apply the spectral flow argument and infer the existence of these charged states in the spectrum.

Given the conformal dimensions above, the correlator in eq. \ref{3pointchi} effectively reduces to a product of a two-point function and a one-point function $\langle \psi \psi \rangle \langle\chi(z,\bar{z}) \rangle$.
Furthermore, since we had assumed a unitary internal CFT, this forces the one-point function to vanish, $\langle\chi\rangle=0$, thus killing the entire factor.

\medskip
We now have all the pieces to combine in the long-range forces between two charged states. We have the resultant of the forces
\begin{equation}\label{resultant}
	F = - \frac{4k_D^2}{\alpha' mm'} \Big( \frac{\alpha'}{2}mm' - \delta_{ab} Q^a Q^b  \Big) \Big( \frac{\alpha'}{2}mm' - \delta_{\tilde{a}\tilde{b}} \tilde{Q}^{\tilde{a}}\tilde{Q}^{\tilde{b}} \Big).
	\end{equation}
    Recall that the spectrum we found in eq. \ref{spectrum} gave us the lightest states of a given charge. To these we can still add external oscillators to obtain massive states with a spectrum
    \begin{equation}
    \frac{\alpha'}{4} m^2 = \frac{1}{2} \max(Q^2, \tilde{Q}^2) +N -1.\end{equation}
Setting the masses and charges equal in eq. \ref{resultant}, we find, in table \ref{tableselfforce}, that the self-force ${F}_{\text{self}} $ exhibits different behaviors for the different $N$ oscillators of a given charge.		
\begin{table}\centering\begin{tabular}{cl}
	$N=0$	& $ {F}_{\text{self}} >0 $: if	$|Q^2 - \tilde{Q}^2| >2 $\\
	$N=0$	& ${F}_{\text{self}}  =0 :$ if 	$|Q^2 - \tilde{Q}^2| =2 $\\
	$N=0$	& ${F}_{\text{self}}  <0 $ if	$|Q^2 - \tilde{Q}^2| = 0 :$\\
		${N=1}$	&{$ {F}_{\text{self}} =0 $: always }\\
			$N=2$	& $ {F}_{\text{self}}  <0 $ always 
		\end{tabular}\caption{Self-repulsive force in a tower of states of charge $(Q,\tilde{Q})$.}\label{tableselfforce}\end{table}
Luckily the $N=1$ states are always self-repulsive\footnote{Recall that saturating the inequality is also a valid solution for the RFC.}.
Thus we have found that an entire sublattice of the charge lattice (the period lattice) contains self-repulsive states and furthermore these are universal as they originate from the spectral flow argument from the graviton (the identity in the internal CFT).

\subsection{Black hole solution and WGC from RFC}
 Up to to this point, we have found universal states in all string theories, since they are connected to graviton states and, by connecting EFT to worldsheet correlators, derived that (excitations of) these states satisfy the RFC.
 Now, we need to perform the last step showing that the existence of these states implies that there are superextremal states in the spectrum.
	
A spherically symmetric black hole solution is of the form \cite{Harlow:2022ich}
\begin{align}
	d s^2 &=  - e^{2 \psi(r)} f (r) d t^2 +
	e^{- \frac{2}{d - 3} \psi(r)}  \left[ \frac{d r^2}{f (r)} + r^2 d
	\Omega^2_{d - 2} \right] \nonumber\\ 
	F^a_2 &= \frac{\mathfrak{f}^{a b} (\phi(r)) Q_b}{V_{d - 2}}  \frac{e^{2\psi(r)}}{r^{d - 2}} d t \wedge d r \,,  \quad 
	 f (r) = 1 - \frac{r_h^{d - 3}}{r^{d - 3}}
\end{align}
and the mass of these solutions can be written as a functional
\begin{equation}\label{massufunctional}
M_{\text{BH}} . = 
	\frac{1}{2}  \int_0^{z_h} e^{2 \psi}  \bigl[\underbrace{\mathfrak{f}^{a b} Q_a
		Q_b - k_N M (\phi)^2 - G^{i j} M_{, i}  M_{, j} }_\star\bigr] d z -
	f e^{\psi} M(\phi) \biggr|_0^{z_h}
\end{equation}
of the arbitrary function $M(\phi)$.
If $M(\phi)$ is chosen such that $\star\geq 0$, then this provides a lower bound on the mass of a black hole solution of given charge $Q$,
\begin{equation}
M_{\text{BH}}\geq f e^{\psi} M(\phi) \biggr|_0^{z_h} = M(\phi_\infty)\equiv M_{EXT}(Q).
\end{equation}

\medskip
	Now suppose there is a self-repulsive particle, i.e.
	\begin{equation}\label{RFC2}
	\mathfrak{f}^{a b} q_a q_b - k_N m^2 - G^{i j} m_{, i}  m_{, j} \geq 0\footnote{This is the same as the RFC bound \ref{RFC}.}.
	\end{equation}
	Then, taking $M(\phi) = \Lambda m(\phi)$ $\implies$ $M_{\text{BH}} \geq \Lambda m(\phi)$.
    This may be thought heuristically as describing such a black hole with charge $Q=\lambda q$ as a condensate of self-repulsive states satisfying eq. \ref{RFC2}.
    Note that this also serves the purpose of stating the WGC as a limit for $\lambda\to \infty$ of the statement 
	$\lambda m \leq M_{\text{EXT}}(Q= \lambda q)$. The limit automatically enforces that we are only considering large (much heavier than the Planck scale) black holes.

In the previous section we had in fact found that the first excited states ($N=1$) satisfy the RFC for any $Q$.
Since they have vanishing self-force, these states satisfy
\begin{equation}
\frac{Q_{\text{BH}}}{M_{\text{BH}}} \leq \frac{Q}{M(\phi)} =  \frac{\Lambda q}{\Lambda m(\phi)} =  \frac{q}{m} 
\end{equation}
saturating the WGC at tree level.
However, we have no control over higher order corrections, these may spoil the equality in the direction violating of WGC. Regardless, since in bosonic string theory we do not have BPS conditions, we do not expect that any states would saturate the WGC bound.
However, we do have a saving grace. The states that we used in the above expressions were the $N=1$ excitations of a tower of a given charge.
Therefore, we also have at our disposal the $N=0$ states. As shown in the above table \ref{tableselfforce}, these do not have a fixed (sign of the) self-force, so they do not all satisfy the RFC. Nonetheless, they are parametrically, as $1/\alpha'$, lighter than the $N=1$ states and so, since the $N=1$ states saturate the WGC bound at tree level, the $N=0$ states are superextremal for any $(Q,\tilde Q)$ in the period lattice.
Thus, the end result is that with the caveats of our derivation, the Ooguri-Vafa version of the sublattice WGC is satisfied in all bosonic string theories.
    
\section{Outlook}\
The work we presented here is a summary of \cite{Heidenreich:2024dmr} which provides a proof of the weak gravity conjecture in perturbative bosonic string theory at tree level.
In particular, we showed that the Ooguri-Vafa version of the sublattice WGC holds, finding a sublattice of the full charge lattice of the theory occupied by superextremal states satisfying the WGC bound which are parametrically lighter than states that would saturate the condition.
Many of the ingredients had previously appeared in the literature, however the explicit computation of the long-range forces for a generic theory, and a comprehensive general top-down approach is novel.  

Though this may be seen as only a proof of concept for the approach that we take, we expect that the generalization to the superstring case should have the same salient features that we described here.
In addition to avoiding pathologies due to tachyons, in a superstring setting the spectrum will be more complicated, possibly containing both exactly extremal states and strictly superextremal ones. According to the Ooguri-Vafa criterion the former have to be BPS, while the latter cannot. This is something we have to check for consistency in the superstring derivation.
Furthermore, the various sectors of the superstring spectrum are mapped into each other by the modular group.
Although the NSNS sector may appear, intuitively, very similar to what we found in the bosonic analysis, the interplay between conditions on the spectrum and modular transformations is a fundamental detail that needs to be addressed \cite{HL26}.

The limitation to constraining the spectrum at tree level, coming from the fact that we have used modular invariance and the spectral flow on the one-loop partition function may not be as severe as it might appear. On one hand, it would be a nice result if we could repeat the argument exploiting the modular group of higher genus surfaces. On the other hand, since we found a parametric separation between the WGC-satisfying charged states and the bound saturation, we expect that higher loop corrections will not spoil the result.

A key step in connecting the RFC to the WGC using black hole solutions  (eq. \ref{massufunctional}) is the requirement that the self-repulsive condition needs to hold everywhere in moduli space, or more precisely in the entire region that such a black hole can explore, otherwise we could not impose a general bound on the mass of a given black hole.
This naturally raises the alarm in the case that we ventured in regions of moduli space at strong coupling. If this happened, we could not trust the black hole solutions and thus could not compare the spectrum to anything. It turns out that, starting with the perturbative solutions that we used, we can show (with mild assumptions) \cite{HL26ext} that these stay within a region of moduli space that is weakly coupled, therefore we can ensure the implication of the WGC from the RFC for any such black hole solution.

Finally, the WGC as formulated and analyzed here is for flat space-time. There are also formulations and checks in AdS space \cite{Lin:2025wfe}, in particular exploiting the AdS/CFT correspondence \cite{Nakayama:2015hga, Cho:2023koe}. While we may not have an analogous worldsheet description in this scenario, it would be interesting to study if and how much of the approach we used could be applied to this case.
In addition, similar worldsheet techniques to what we described here could be applied to shed light on other swampland conjectures \cite{Aoufia:2026bau, Basile:2026nln, Ooguri:2024ofs}
 and/or even unify them.

\subsection*{Acknowledgments}

It is a pleasure B. Heidenreich for collaborations on which this review is based. I would also like to thank I. Basile and G. Casas for helpful  comments on the manuscript and discussions. 
This work was supported by the Ayuda RYC2023-043268-I funded by MICIU/AEI/10.13039/501100011033 and FSE+, as well as through the grants PID2024-156043N B-I00, PID2021-123017NB-I00 and CEX2020-001007-S, funded by MCIN/AEI/10.13039/50110 0011033, and ERDF, EU.

\end{document}